\documentclass[12pt,letterpaper]{article}

\usepackage[includeheadfoot,
            marginratio={1:1,2:3}, 
            width=412pt, 
            height=688pt,]{geometry}

\usepackage{amsmath}
\usepackage{amsfonts}
\usepackage{amssymb}
\usepackage{stmaryrd}
\usepackage{graphicx}

\usepackage{empheq}
\usepackage{paralist}

\newcommand{\nc}{\newcommand}
\nc{\lb}{\llbracket}
\nc{\rb}{\rrbracket}
\nc{\gl}{\llbracket}
\nc{\gr}{\rrbracket}
\newcommand{\tri}{\hspace{-3.5pt}\vartriangle\hspace{-3.5pt}}
\newcommand{\blacktri}{\blacktriangle}

\newcommand{\eq}[1]{\begin{equation}
                     \begin{split} #1 \end{split}
                     \end{equation}}

\newcommand{\ov}{\overline}

\allowdisplaybreaks[2]
\numberwithin{equation}{section}

\begin{document}

\vspace*{-1.5cm}
\begin{flushright}
  {\small
  MPP-2013-307\\
  LMU-ASC 80/13\\
  }
\end{flushright}

\vspace{1.5cm}
\begin{center}
{\LARGE
Non-associative Deformations of Geometry \\[0.2cm]
  in Double Field Theory \\[0.2cm]
}
\vspace{0.4cm}

\end{center}

\vspace{0.35cm}
\begin{center}
  Ralph Blumenhagen$^{1}$, Michael Fuchs$^{1}$, Falk Ha\ss ler$^{2}$, \\[0.1cm]
Dieter L\"ust$^{1,2}$ and Rui Sun$^{1}$
\end{center}

\vspace{0.1cm}
\begin{center} 
\emph{$^{1}$ Max-Planck-Institut f\"ur Physik (Werner-Heisenberg-Institut), \\ 
   F\"ohringer Ring 6,  80805 M\"unchen, Germany } \\[0.1cm] 
\vspace{0.25cm} 
\emph{$^{2}$ Arnold Sommerfeld Center for Theoretical Physics,\\ 
               LMU, Theresienstr.~37, 80333 M\"unchen, Germany}\\

\vspace{0.2cm}

 \vspace{0.3cm} 
\end{center} 

\vspace{1cm}


\begin{abstract}
Non-geometric string backgrounds were proposed to be related to 
a non-associative deformation of the space-time geometry.
In the flux formulation of
double field theory (DFT),  the structure  of 
mathematically possible 
non-associative deformations is analyzed in detail.
It is argued that on-shell there should not be any violation
of associativity in the effective DFT action.
For imposing either the strong or the weaker closure constraint
we discuss two possible non-associative deformations
of DFT featuring two different ways how on-shell
associativity can still be kept.
\end{abstract}

\clearpage




\section{Introduction}
\label{sec:intro}

In string theory, a large scale geometric target space is rather an
emergent phenomenon. The basic starting point  is the two-dimensional
field theory on the world-volume of the probe string equipped with
the fundamental paradigm that  on-shell solutions
of string theory are provided  
by two-dimensional conformal field theories (CFTs)
with  the critical central charge. However, 
the generic left-right asymmetric CFT does not correspond 
to a  fixed point of a non-linear
sigma model with a geometric target space.
Since  string theory is strongly believed to provide a consistent theory of 
quantum gravity, one may wonder to which  non-geometric
generalizations of the target space-time the generic asymmetric
CFT corresponds to. This could also enlighten relations 
to complementary  target-space approaches to quantum gravity,
like loop quantum gravity 
or non-commutative geometry.
 
During the last years some progress has been made towards
a better understanding of this non-geometric regime of string theory.
In fact, the recent developments
go precisely in the direction of providing a quasi-geometric 
description of these asymmetric conformal field theories.
T-duality is a left-right asymmetric transformation, so
that it served as the main tool to shed some light into this
mainly unexplored regime of the string theory landscape.

In \cite{Shelton:2005cf} the simple closed string background
of a flat space with constant $H$-flux and dilaton was considered. 
Successively applying the Buscher rules, one gets
the well-known chain of T-dual configurations
\eq{
  H_{ijk} \;\xleftrightarrow{\;\; T_{k}\;\;}\;
   F_{ij}{}^{k} \;\xleftrightarrow{\;\; T_{j}\;\;}\;
  Q_{i}{}^{jk} \;\xleftrightarrow{\;\; T_{i}\;\;}\;
  R^{ijk} \; .
}
The last two were  argued to be non-geometric. The $Q$-flux case
 is still geometric locally but the transition functions
involve non-geometric T-duality transformations, whereas
the $R$-flux case is considered to be even locally non-geometric.

A simple argument shows that this background does not allow the notion
of a point \cite{Wecht:2007wu}. 
Let us repeat it here to make clear that something drastic
must happen for these backgrounds. Consider a D3-brane wrapping 
a three-torus carrying  a constant three-form $H$-flux.
In fact such a configuration is not allowed as it suffers from the
Freed-Witten anomaly \cite{Freed:1999vc}, i.e. it violates the
Bianchi identity $d{\cal F}=H$ for the gauge flux on the brane.
Now, by formally applying a T-duality
along all three directions of the torus, one gets a D0-brane 
with transverse $R$-flux. Thus, placing  a point-like object
in an $R$-flux background is not allowed. This suggests that
one has an uncertainty relation like 
$\Delta x\,\Delta y\,\Delta z\ge  \ell_s^4\, R^{xyz}$,  pointing
towards a relation to non-commutative geometry.
 
Indeed, it was abstractly argued that the $R$-flux  involves a 
non-associativity of the coordinates \cite{Bouwknegt:2004ap}.
More recently it was found 
\cite{Blumenhagen:2010hj,Lust:2010iy,Blumenhagen:2011ph,Condeescu:2012sp,Andriot:2012vb} by explicit string and CFT computations
that the string geometry indeed becomes non-commutative and non-associative
for closed strings that are winding and moving in non-geometric backgrounds.
Concretely, the equal-time cyclic double-commutator of 
three local coordinates was  found to be  
\eq{
  \label{result2}
   \bigl[x^i,x^j,x^k \bigr] 
   &=  \begin{cases} 0 & H{\rm -flux} \\
                \ell_s^4\,R^{ijk}  & R{\rm -flux}
     \end{cases}\; .
}
The same result arises from  a commutator algebra 
\eq{
\label{commalg}
           [x^i,x^j]= \frac{i}{ 3\hbar} \ell_s^4\, R^{ijk}\, p_k\, , \qquad [x^i,p_j]=i \hbar\, \delta^i{}_j
} 
so that the Jacobiator  gives precisely \eqref{result2}.
If also $Q$-flux is present the commutator was argued to be generalized to
\eq{
\label{commalggen}
           [x^i,x^j]= \frac{i}{ 3\hbar} \ell_s^4\, \Big(R^{ijk}\, p_k+
     Q_k{}^{ij} w^k\Big)\, ,
} 
where $w^k$ is the winding operator.
Analogous  relations were also derived in the framework of matrix theory 
in \cite{Chatzistavrakidis:2012qj}.

In \cite{Blumenhagen:2011ph} this background was investigated 
using conformal
perturbation theory and, analogous to the open string story \cite{Seiberg:1999vs},
on-shell string scattering amplitudes of  tachyons were computed.
Actually, for both constant $H$-flux and $R$-flux the final scattering amplitude
was associative, as expected from crossing symmetry of conformal correlation
functions. However, prior to invoking momentum conservation, 
there was a difference between the $H$- and $R$-flux case, namely
 the appearance of  world-sheet independent phase factors.
For the $H$-flux the holomorphic and anti-holomorphic phases directly
canceled each other while for the $R$-flux they added up.
These phases could be encoded (at least at linear order in $R^{ijk}$) 
in the tri-product\footnote{Choosing $f=\exp(i p_1 x)$ and similar for $g,h$ the momentum conservation
can be implemented by integrating the tri-product \eqref{threebracketcon}, 
so that the
order $\ell_s^4$ correction becomes
\eq{
        \int d^nx \, R^{ijk}\, p^1_i\, p^2_j\, p^3_k\,
        e^{i(p^1+p^2+p^3)\cdot x}=
           R^{ijk} \, p^1_i\, p^2_j\, p^3_k\, \delta(p^1+p^2+p^3)=0\, .
}
The aim of this paper is to generalize this result to non-constant
fluxes on a curved space.
}
\eq{
\label{threebracketcon}
   (f\,\tri\, g\, \tri\, h)(x) = \exp\biggl(
   {\ell_s^4\over 6}\, R^{ijk}\,
      \partial^{x_1}_{i}\,\partial^{x_2}_{j}\,\partial^{x_3}_{k} \biggr)\, f(x_1)\, g(x_2)\,
   h(x_3)\Bigr|_{x} \;.
}
The three-bracket can then be defined as
\begin{equation}
\label{antisymtripcon}
   \bigl [x^i,x^j,x^k \bigr]=\sum_{\sigma\in S_3} {\rm sign}(\sigma) \;  
     x^{\sigma(i)}\, \tri\,  x^{\sigma(j)}\, \tri\,  x^{\sigma(k)} \; ,
\end{equation}
where $S_3$ denotes the permutation group of three elements.
Note that, formally  one could also define 
such a tri-product with $H^{ijk}$ instead of $R^{ijk}$. The tri-product
\eqref{threebracketcon} as well as an associated momentum dependent star-product
was also derived in \cite{Mylonas:2012pg,Bakas:2013jwa} 
by starting with  the non-associative 
commutator algebra \eqref{commalg}. In addition the non-commutative and non-associative phase space structure of DFT as well as the magnetic
field analogue of the string $R$-flux model
was discussed in \cite{Bakas:2013jwa}.

Besides these example based arguments, there was a successful
approach to develop a manifestly T-duality, i.e. $O(D,D)$, covariant
formulation of the dynamics of the massless modes of string theory.
This was initiated in \cite{Siegel:1993xq,Siegel:1993th}
and pushed forward more recently in \cite{Hull:2009mi,Hull:2009zb,Hohm:2010jy,Hohm:2010pp}.
In this so-called double field theory (DFT) framework 
(see \cite{Zwiebach:2011rg, Aldazabal:2013sca,Hohm:2013bwa} for  reviews)
one doubles the number of target space coordinates
by also introducing winding coordinates. 
It turned out that
this is a constrained theory, where usually the weak and
the strong constraint are imposed. 
Then, locally one ends up on a $D$-dimensional 
slice of the $2D$-dimensional doubled geometry, which can be rotated to
the supergravity frame via an $O(D,D)$ transformation.

DFT  is related to
generalized geometry \cite{Hitchin:2004ut,Gualtieri:2003dx, Grana:2008yw} 
by setting the winding coordinates
to zero while keeping the doubled  tangent bundle ${\rm TM}\oplus {\rm TM}^*$.
Moreover, it admits all the local
symmetries, usual and winding diffeomorphisms,  
to allow for a global description of, for instance,
the $Q$-flux and $R$-flux backgrounds. 
This is possible as T-duality exchanges ordinary and winding coordinates
so that for these non-geometric backgrounds there appears a
winding coordinate dependence either in the transition functions 
between two charts ($Q$-flux) or in the definition of the 
flux itself ($R$-flux). Thus, non-geometry just means explicit winding
coordinate dependence in the background fluxes or in the transition functions.

There exist essentially two formulations of DFT. First, there is 
the generalized metric formulation, which was developed in a series
of papers \cite{Hull:2009mi,Hull:2009zb,Hohm:2010jy,Hohm:2010pp}. 
Here one invokes the so-called strong constraint to guarantee
e.g. closure of the symmetry algebra (the C-bracket).
Based on the previous work 
\cite{Siegel:1993xq,Siegel:1993th,Hohm:2010xe}  
and \cite{Hohm:2011ex,Aldazabal:2011nj,Geissbuhler:2011mx,Grana:2012rr}, 
in \cite{Geissbuhler:2013uka} a second formulation of DFT has been provided
which from the onset incorporates the relation to gauged supergravity
theories. This is the so-called flux formulation of DFT, which
was shown to be equivalent to the generalized metric formulation,
up to boundary terms and terms vanishing by the strong constraint.
However, as will also be essential for our investigation,
it allows to move away from the strong constraint and admit
truly non-geometric duality orbits of fluxes in the sense of
\cite{Dibitetto:2012rk}. In fact, it makes use of the observation that
requiring  only closure of the symmetry algebra provides a  
weaker constraint than the strong constraint. 
A weakening of the strong constraint was first discussed in
\cite{Hohm:2011cp}.
Maybe the simplest examples are given
by Scherk-Schwarz reductions \cite{Scherk:1978ta,Scherk:1979zr} of DFT
\cite{Aldazabal:2011nj,Geissbuhler:2011mx,Grana:2012rr, Geissbuhler:2013uka}
(see also \cite{Berman:2012uy,Berman:2013cli}). Note that  
in \cite{Condeescu:2013yma}
concrete examples of asymmetric orbifold CFTs were presented for which
 evidence was provided that they do correspond to such
non-geometric duality orbits.

It was observed that, in DFT, which is a priori a background independent formalism, generalized coordinate transformations compose in a non-standard
manner, such that the composition is non-associative \cite{Hohm:2012gk}. 
However this non-associativity vanishes after imposing the strong constraint
on arbitrary fields. Besides that, in DFT
no notion 
of a non-associative, background dependent deformation of the geometry
is visible. Hence it is puzzling how DFT can be reconciled with
the aforementioned 
claim  that the $R$-flux is related to such a non-associative deformation,
as described for constant flux via the tri-product \eqref{threebracketcon}.
The resolution of  this paradox is the purpose of this paper.
To this end, we identify two important aspects: 
\begin{itemize}
\item{
First, as is apparent from \eqref{threebracketcon}, the non-associativity 
is claimed to arise for an $R$-flux background 
contracted with ordinary partial derivatives  $\partial/\partial x^i$.
Note that, in this sense, the DFT T-dual of the $H$-flux background 
on ordinary space
is an $R$-flux background on winding space.}
\item{
Second, in quantum theories, where observables are operators 
acting on some Hilbert space, one can get non-commutativity, 
but the product of operators is always associative.
Since conformal field theories are ordinary (2-dimensional) quantum theories,
on-shell, i.e. if  the string equations of motion are satisfied,
there should better not be any violation of associativity in CFT on-shell scattering amplitudes.}
\end{itemize}

\noindent
Indeed, in conformal field theory one requires crossing symmetry of the
operator product expansion, which is related to the Jacobi identities
for the algebra of the modes of the conformal fields.  
In string theory, from   on-shell scattering amplitudes,
one can determine an effective theory for the massless modes, which
by construction does not show any on-shell sign of non-associativity.
Therefore, we conclude that any admissible non-associative deformation
given by  a non-associative tri-product like \eqref{threebracketcon}
should have a trivial effect on the effective field theory, when going on-shell.
However it is a priori not clear whether the off-shell effective string action
is sensitive against non-associative deformations of the underlying geometry.

As we will discuss, the main result of this paper is that, on the level
of the effective action, a non-associative deformation of  the DFT 
generalization of both the $H$-flux and the $R$-flux only leads at most
to boundary terms. For the first one has to invoke the DFT equations
of motion, whereas the second deformation turns out to be 
trivial once one imposes either the strong or even the closure 
constraint.

A similar reasoning also applies to the case of open strings
ending on D-branes supporting a non-trivial, in general non-constant
gauge flux. The case when this product becomes non-associative
was analyzed in a series of 
papers \cite{Cornalba:2001sm,Herbst:2001ai,Herbst:2003we}. 
Thus, before we move on
to briefly review  the flux formulation of DFT in section \ref{sec_DFT},
we review in section \ref{sec_two} two known examples of non-associativity, namely
the system of an electric charge moving in a magnetic monopole field and
a D-brane carrying non-constant gauge flux.
In section \ref{sec_main} we will  analyze possible tri-products
for DFT. As we will see, a priori  there are two candidates, one related 
to the tri-product \eqref{threebracketcon} with  $H$-flux  and one
to the tri-product with $R$-flux. Both cases will be discussed in detail.

\section{Non-associativity in physics}
\label{sec_two}

In this section we review two instances where a non-associative 
structure has appeared in physics. First, we recall the story of
quantizing the motion of an electrically charged particle in a magnetic field.
Second, the effective theory on a D-brane with non-constant 
magnetic background field turned on is considered.
This gives a non-vanishing $H=dB$ flux, which in general leads
to a non-associative star-product. 

\subsection{Non-associativity for magnetic monopoles}
\label{sec_magnetic}

As it is known for some time \cite{Jackiw:1984rd,Jackiw:1985hq,Wu:1984wr,Grossman:1984fs,Grossman:1985hz},
non-associativity emerges when considering the quantization of a charged particle in the background of a magnetic monopole.
Hence in this context immediately the question arises how the apparent emergence of non-associativity can be reconciled with the basic principles of quantum mechanics, where associativity of
all operators is mandatory. This issue was recently addressed in
\cite{Bakas:2013jwa}, where also some remaining puzzles of the earlier work
were resolved.

Here, let us just recall a few facts about this system following essentially
\cite{Jackiw:1984rd,Jackiw:1985hq}.
The commutator algebra between position and momentum of a particle in a 
background magnetic field
$\vec{B}$ in three space-dimensions takes the following form
\begin{equation}
[x^i ,  p_j] = i\hbar  \delta^{i}_j ~, ~~~~~
[x^i ,  x^j] = 0 ~, ~~~~~ [p_i ,  p_j] = {i\hbar }\,e \, \epsilon^{ijk} B_k (\vec{x})\, .
\label{gen}
\end{equation}
In turn, the Jacobiator becomes
\begin{equation}
[p_i , p_j , p_k] = - e\hbar^2 \,\epsilon^{ijk}\, \vec{\nabla} \cdot \vec{B} \, 
\label{dirac}
\end{equation}
with   $\vec{\nabla} \cdot \vec{B}=4\pi \rho_m$ in
Gaussian-cgs units.
These relations have  the analogous  form as the commutators  (\ref{commalg}) and three-bracket 
(\ref{result2}) after exchanging the role of momentum and position variables
in these equations.
Now, consider the finite translation operators $U(a)=\exp({i\over \hbar} a\cdot p)$.
Using the Baker-Campbell-Hausdorff formula one obtains
\eq{
      U(a)\, U(b)=\exp\left(-{i\,e\over \hbar}\,  \Phi_{(a,b)}\right)\, U(a+b)
}
where $\Phi_{(a,b)}={1\over 2} (a\times b)^k B_k$ denotes the
magnetic flux through the (infinitesimally small) 
triangle spanned by the two vectors $(a,b)$.
Similarly, one can compute the associator of three $U$s 
\eq{
\label{nonmagn}
     \big( U(a)\, U(b) \big) U(c)=\exp\left(-{i\, e\over \hbar}\,
     \Phi_{(a,b,c)}\right)\, U(a) \big( U(b)\, U(c) \big) 
}
where $\,\Phi_{(a,b,c)}={1\over 6} [(a\times b)\cdot c] \vec\nabla\cdot  \vec B\,$ denotes the magnetic flux through the
tetrahedron spanned by the three vectors $(a,b,c)$. Due to
Gauss law this is nothing else than the magnetic charge $4\pi\, m$
sitting inside the tetrahedron.
Therefore, the non-associativity  \eqref{nonmagn} vanishes
if the phase is trivial, i.e.
\eq{
          {e\, m\over \hbar} = {N\over 2}
}
with an integer $N$. This is Dirac's quantization rule for
the magnetic charge. Thus we can cite from the abstract of
\cite{Jackiw:1984rd} 'Insisting that finite translations be associative leads to Dirac's monopole quantization condition'.

As discussed in \cite{Bakas:2013jwa}, only for the case of the 
magnetic monopole the classical equations of motion of  a charged particle 
are still integrable. In this case, the so-called
Poincar\'e vector provides an integral of motion, and angular momentum is
still preserved. For a continuous  magnetic charge distribution $\rho (x)$
angular symmetry gets broken.

In this paper, we are essentially generalizing the above mentioned logic
by clarifying 
how the non-associative tri-product deformation of 
the DFT action can be made 
consistent with the requirements from CFT scattering amplitudes.
The only main difference is that we are not considering
quantized fluxes and momenta but the case where
these are in general non-rational and space-time dependent.
However, the main message still is that from the requirement
of absence of non-associativity we can learn something very essential 
about the system.

\subsection{Open string with non-associative star product}
\label{sec_open}

Let us recall that the conformal field theory of an open
string ending on a D-brane supporting a non-trivial 
gauge flux ${\cal F}=B+2\pi\alpha' F$ features 
a non-commutative geometry.

Indeed, by computing the disc level scattering amplitude of
$N$-tachyons, certain relative phases appear which 
for constant gauge flux can be described by the Moyal-Weyl
star-product
\eq{
\label{starproduct}
   (f\,\star\, g)(x)= \exp\biggl(
   i\, {\ell_s^2\over 2}\, \theta^{ij}\,
      \partial^{x_1}_{i}\,\partial^{x_2}_{j} \biggr)\, f(x_1)\, g(x_2)\Bigr|_{x} \;,
}
where the relation of the open and closed string quantities is
\eq{
\label{openclosed}
     G^{-1}+\theta=(g+{\cal F})^{-1}\, .
}
In the Seiberg-Witten limit the OPE exactly becomes the 
Moyal-Weyl star-product. This non-trivial product of functions
lead to the non-commu\-tative Moyal-Weyl plane with
$[x^i,x^j]=i\,\ell_s^2\, \theta^{ij}$. 
That in the on-shell string scattering amplitudes
such a non-commutativity can show up, is possible because
the conformal $SL(2,\mathbb R)$ symmetry group only leaves
the cyclic order of the inserted vertex operators invariant.
By the same reason,  the non-commutativity must be such
that, on-shell, it preserves cyclicity.

There is no need to only consider
a constant antisymmetric two-vector $\theta^{ij}$. Indeed, 
in \cite{Kontsevich:1997vb} 
it  has been shown that for every Poisson structure $\theta^{ij}$
one can define a corresponding  associative star-product, which 
will also involve derivatives of the Poisson structure.
The same product can also be considered for a quasi Poisson structure,
but then leads to a non-associative star-product.
This is related to the physical situation of 
 an open string ending on a D-brane
with generic non-constant $B$-field, i.e. non-vanishing
field strength $H$.
At leading order in derivatives this leads to 
a non-commutative product
\eq{
     f\circ g=f\cdot g + 
  &i\, {\ell_s^2\over 2} \theta^{ij}\, \partial_i f\, \partial_j g -
   {\ell_s^4\over 8} \theta^{ij}\theta^{kl}\, \partial_i\partial_k f\, \partial_j\partial_l g  \\
  &-{\ell_s^4\over 12} \big(\theta^{im}\partial_m \theta^{jk}\big) \big(
  \partial_i\partial_j f \, \partial_k g + \partial_i\partial_j g\, \partial_k f\big) \ldots\, .
}
The associator for this product becomes 
\eq{
\label{openasso}
    (f\circ g)\circ h-f\circ (g\circ h)={\ell_s^4\over 6}\,
      \theta^{ijk}\, \partial_i f \,\partial_j g \,\partial_k h+\ldots
}
with $\theta^{ijk}=3\,\theta^{[\underline{i}m}\partial_m
  \theta^{\underline{jk}]}$, which precisely vanishes for a Poisson tensor.
But now the puzzle arises that in the open string CFT we should {\it not} see
the effect of such a non-associative deformation of the underlying 
space-time. Indeed this question was analyzed in some detail 
in \cite{Herbst:2001ai,Herbst:2003we}
and we briefly repeat their essential observation  here.

{}From the open string scattering amplitudes 
one can determine the low-energy effective
action so that also the effect of the non-associativity 
in its  quantum deformation should be trivial.
Indeed, consider the DBI action
\eq{
    S_{\rm DBI}=\int d^nx  \sqrt{g+{\cal F}}
}
and vary it with respect to the gauge potential $A$ in ${\cal F}=B+dA$.
One gets 
\eq{ 
\label{eoma}
 \partial_i\left( \sqrt{g+{\cal F}} \left[ (g+{\cal F})^{-1}\right]^{[\underline{ij}]}\right) =\partial_i\left( \sqrt{g+{\cal F}} \, \theta^{ij}\right)=0
}
where we have used \eqref{openclosed}.
Then, it directly follows that up to leading  order in $\partial \theta$
the $\star$-product satisfies the property
\eq{
\label{intiprop}
     \int d^nx \sqrt{g+{\cal F}}\, f\circ g= \int d^nx \sqrt{g+{\cal F}}\, f\cdot g
\, .}
Indeed, e.g. at order $O(\ell_s^2)$   the difference between the left and the right hand side is
a total derivative on-shell
\eq{
     i\, {\ell_s^2\over 2}   \int d^nx \sqrt{g+{\cal F}}\, \theta^{ij}\, \partial_i f\, \partial_j g
    = i\, {\ell_s^2\over 2} \int d^nx\, \partial_i \!\left(\sqrt{g+{\cal F}}\, \theta^{ij}\, f\, \partial_j g\right) =0\, 
}  
where here and in the following sections we always assume
that the functions $f,g$ are sufficiently well behaving so that
integrals over total derivatives vanish.
Thus, as expected from CFT, in the effective action the 
product of two functions is commutative (cyclic), once
the background satisfies the string equations of motion.

Similarly, the associator below the integral also gives a total
derivative at leading order in $\partial \theta$. 
E.g. at order $O(\ell_s^4)$ we find
\eq{
\label{assoprop}
        \int d^nx \sqrt{g+{\cal F}}\; &\Big( (f\circ g)\circ h-f\circ (g\circ h)\Big)=\\
   &{\ell_s^4\over 6}
\int d^nx\, \partial_i \!\left(\sqrt{g+{\cal F}}\, \theta^{ijk}\, f\, \partial_j g\, \partial_k h\right) =0\, ,
} 
where we have used 
\eq{
                 \partial_i\Big( \sqrt{g+{\cal F}} \, \theta^{ijk}\Big)=0\, ,
}
which can be seen by expanding $\theta^{ijk}$ and successively employing
the equation of motion \eqref{eoma} and the anti-symmetry of $\theta^{ij}$.
The two relations  \eqref{intiprop} and \eqref{assoprop} also hold for higher orders
in derivatives of $\theta^{ij}$ \cite{Herbst:2003we}. Note, that as here one is using the 
DBI action, the star-product is exact in $\alpha'$ at leading order
in $\partial\theta$. 
Thus, we conclude that, as expected from the open string conformal
field theory, on-shell the  non-associativity 
of the $\circ$-product is not visible.

In the following we will generalize this kind of analysis to the
closed string case. Since there we are dealing with non-geometric
fluxes, the appropriate  framework to discuss it is double field
theory.
Therefore, let us recall those aspects of DFT which will be used
in the main section \ref{sec_main}.

\section{Flux formulation of DFT}
\label{sec_DFT}

In this section we summarize the main features of the flux formulation
of DFT, as it has been described 
in \cite{Geissbuhler:2013uka,Aldazabal:2013sca}, based on the earlier
work  \cite{Siegel:1993xq,Siegel:1993th} and \cite{Aldazabal:2011nj,Geissbuhler:2011mx,Grana:2012rr}. 
For more details we refer to these  papers.

\subsection{Basics of DFT}

The main new feature of DFT is that
one doubles the number of coordinates
by introducing   winding coordinates $\tilde x_m$ and arranges
them into a doubled vector $X^M=(\tilde x_m,x^m)$.
One defines an $O(D,D)$ invariant metric
\eq{
 \eta_{MN}=
    \left(\begin{matrix}  0 &  \delta^m{}_n \\
            \delta_m{}^n & 0 \end{matrix}\right) \, 
}
and introduces a generalized bein $E^A{}_M$ with metric
\eq{
            S_{AB}=
 \left(\begin{matrix}  s^{ab} &  0 \\
            0 & s_{ab} \end{matrix}\right) \, 
}
with $s_{ab}$ being the flat $D$-dimensional Minkowski metric.
The most generic parameterization of this generalized bein reads
\eq{
\label{gennonhol2}
       E^A{}_M=
 \left(\begin{matrix}  e_a{}^m &  e_a{}^k\, B_{km} \\
            e^a{}_k \beta^{km}  & e^a{}_m+e^a{}_k \beta^{kl} B_{lm}  \end{matrix}\right) \, ,
}
with the ordinary bein $e_a{}^m s^{ab} e_b{}^n=G^{mn}$.
Note that \eqref{gennonhol2}  contains both a  two form $B_{mn}$ and a 
two-vector $\beta^{mn}$.
The flat derivative is defined as
\eq{
\label{deriv}
              {\cal D}^A=E^A{}_M\, \partial^M\, .
}
Using  these beins, one defines the generalized fluxes $F_{ABC}$ as
\eq{
\label{dftflux}
      {\cal F}_{ABC}=3\Omega_{[ABC]}
}
in terms of  the generalized Weitzenb\"ock connection\footnote{For
a recent discussion of the role of a  Weitzenb\"ock connection in DFT 
see \cite{Berman:2013uda}.} 
\eq{
\label{dftomega}
      \Omega_{ABC}={\cal D}_A E_B{}^M\, E_{CM}\, .
}
The components of these DFT fluxes ${\cal F}_{ABC}$ are precisely
the geometric and non-geometric fluxes $H,F,Q$ and $R$
\eq{
     {\cal F}_{abc}=H_{abc}\, , \quad {\cal F}^a{}_{bc}=F^a{}_{bc}\, , \quad 
   {\cal F}_c{}^{ab}=Q_c{}^{ab}\, , \quad  {\cal F}^{abc}=R^{abc}\, .
}
The explicit form of these fluxes in terms of $B$ and $\beta$ 
can be found in 
\cite{Aldazabal:2011nj, Geissbuhler:2013uka, Blumenhagen:2013hva} (see also \cite{Andriot:2012an}). 
For later use we just list the
fluxes for the choice $B_{mn}=0$ in \eqref{gennonhol2}.
Defining 
\eq{
         f^c{}_{ab}= e_i{}^c \Big( \partial_a e_b{}^i - \partial_b
         e_a{}^i\Big)\,, \quad\qquad
\tilde f_a{}^{bc}= e_a{}^i\,  \Big( \tilde\partial^b e_i{}^c -
               \tilde\partial^c e_i{}^b \Big)\; ,
}
one finds $H_{abc}=0$ and the geometric flux $F^c{}_{ab}= f^c{}_{ab}$. 
The non-geometric fluxes are
\eq{
\label{exqflux}
            Q_c{}^{ab}= &\tilde f_c{}^{ab} + {\partial}_c \beta^{ab} + {f}^a{}_{cm} \beta^{mb}
               + {f}^b{}_{cm}\beta^{am}  
}
and 
\eq{
\label{fluxr}
  R^{abc}= &3\Big(\tilde\partial^{[\underline{a}} \beta^{\underline{bc}]}
                    + \tilde f_m{}^{[\underline{ab}}\,
                    \beta^{\underline{c}]m}\Big)+
               3\Big( \beta^{[\underline{a}m} {\partial}_m \beta^{\underline{bc}]}
             +\beta^{[\underline{a}m} \beta^{\underline{b}n}
               {f}^{\underline{c}]}{}_{mn}\Big)\, .
}
Similar to the open string case \eqref{openasso}, 
 the contribution $R^{abc}_{\rm cl}= 3\big( \beta^{[\underline{a}m}
  {\partial}_m \beta^{\underline{bc}]}+\ldots\big)$ can be considered
as the defect for associativity, when we consider $\beta^{ab}$ as
a classical (quasi-) Poisson tensor.

Next, one introduces the T-duality invariant dilaton 
\eq{
e^{-2d}=e^{-2\phi}\sqrt{g}
}
which is used to also define
\eq{
     {\cal F}_A=\Omega^B{}_{BA}+2 E_A{}^M \partial_M d\, .
}
DFT is required to be invariant under a large symmetry group. 
First it is invariant under global $G=O(D,D)$ transformation 
and second it is invariant under a local $H\subset G$ 
symmetry with $H=O(D)\times O(D)$.
This local symmetry acts on the bein as
\eq{
\label{localsymme}
       \delta_\Lambda E_A{}^M = \Lambda_A{}^B \, E_B{}^M\, \qquad {\rm with}\quad
     \Lambda_A{}^C S_{CD}\,  \Lambda_B{}^D = S_{AB}\, 
}
so that they can be viewed as local double  Lorentz transformations.
Besides that, the usual diffeomorphism symmetry is enhanced to
so-called generalized diffeomorphism with infinitesimal parameter
$\xi^M=(\tilde\lambda_m, \lambda^m)$ and generalized
Lie-derivative,  acting e.g. on a doubled vector $V$ as
\eq{
    {\cal L}_\xi V^M= \xi^N \partial_N V^M + (\partial^M \xi_N-\partial_N
    \xi^M) V^N \, .
}
For instance the beins $E_A$ transform as a doubled vector, whereas
the dilaton $d$ transforms as a scalar  density
\eq{
   \delta_\xi d= 
  {\cal L}_\xi d=   \xi^M \partial_M d -{1\over 2} \partial_M \xi^M\, .
}
This allows to define a generalized tensor calculus by defining that
the variation of a tensor with respect to generalized diffeomorphisms
is
\eq{
        \delta_\xi T^{M_1\ldots M_k}={\cal L}_\xi T^{M_1\ldots M_k}\, .
}

In contrast to the usual Lie-derivative, the Lie-derivative 
of a generalized tensor is not automatically again a generalized
tensor. To ensure this, one has to impose the so-called 
{\it closure} constraint
\eq{
\label{closureco}
       \Delta_{\xi_1} ({\cal L}_{\xi_2} T^{M_1\ldots M_k})=0
}
with the anomalous variation $\Delta(\,\cdot\,)=\delta_{\xi}(\,\cdot\,)-{\cal L}_{\xi}(\,\cdot\,)$.

The invariant action of the  flux formulation  of  DFT reads
\eq{
 \label{dftactionfluxb}
   S_{{\rm DFT}}=
\int
&dX \; e^{-2d}\, \bigg[ {\cal F}_A {\cal F}_{A'} {S}^{AA'} +\\
&{\cal F}_{ABC} {\cal F}_{A'B'C'}\,
\Big( \frac{1}{4} {S}^{AA'} \eta^{BB'} \eta^{CC'} 
-\frac{1}{12} {S}^{AA'} {S}^{BB'} {S}^{CC'} \Big)\\
& - {1\over 6}{\cal F}_{ABC}\, {\cal F}^{ABC}- {\cal F}_A {\cal F}^A
\bigg]\, .
}
Note that in CFT we can assign a  world-sheet parity $\Omega$ to every field
(see e.g. \cite{Blumenhagen:2013hva}).  
Then, the terms in the first two lines are $\Omega$-even and
the term in the last line  are $\Omega$-odd.
The DFT action  has to be supplemented by one of the following constraints. 
\begin{itemize}
\item{\underbar{Strong constraint}:
In this case one requires the so-called  weak and   strong constraint
\eq{
  \label{strong_c}
   \partial_M\partial^M=0\,, \qquad
    \partial_M f\, \partial^M g =  {\cal D}_A f \, {\cal D}^A g =0\, 
}
with $f,g$ being the fundamental objects like  $E^A{}_M$ and $\xi^M$.
Locally, up to an $O(D,D)$ transformation these constraints
remove the winding dependence. In particular, the constraints
guarantee the closure constraint. In the following, we  always implement
the weak and strong constraint  for the uncompactified directions.}

\item{\underbar{Closure constraint}: For compact spaces one
can  weaken the strong constraint and only require  that  
 the symmetry algebra closes \cite{Grana:2012rr}, i.e.
that a Lie-derivative of a generalized tensor is again a generalized tensor
\eqref{closureco}.
Scherk-Schwarz reductions are prototype examples, whose
reduced action  is closely related to gauged supergravity
and whose internal spaces are truly non-geometric in the
sense that fields depend on doubled coordinates 
$(y^m,\tilde y_m)$.
}
\end{itemize}

Let us analyze some of the consequences of just imposing the
closure constraint.
First, if $f$ is a generalized scalar, then we can write
\eq{
     {\cal D}_A f=E_A{}^M \partial_M f={\cal L}_{E_A}(f)
}
which by the closure constraint implies that $\Delta_{\xi}({\cal L}_{E_A}
f)=0$. Therefore, ${\cal D}_A f$ is also generalized scalar.
Now, by direct computation one obtains
\eq{
\label{derivvario}
   \Delta_{\xi}({\cal D}_B f)&=\delta_{\xi}({\cal D}_B f)-{\cal L}_{\xi}({\cal
     D}_B f)\\
&=\left({\cal D}^C \xi^M\right) E_{BM}\, {\cal D}_C f=0
} 
Thus, choosing $\xi=E_A$ we can conclude
\eq{
\label{importanta}
\left({\cal D}^C E_A{}^M\right) E_{BM}\, {\cal D}_C f
=\Omega^C{}_{AB}\, {\cal D}_C f=0\, .
}
For a generalized scalar  $g$,
we can also choose $\xi=E_B g$ in \eqref{derivvario}
and, using the relation \eqref{importanta}, obtain 
\eq{
\label{SCfromClosure}
\delta_{AB} \,\, {\cal D}^C g \,\,\,{\cal D}_C f = 0\, .
}
Thus, we conclude that the closure constraint implies that
for scalars $f$ and $g$ the   strong constraint still has
to hold. A particular example which we will use later is
\eq{
\label{importantb}
\left( {\cal D}_C {\cal F}_A\right)  {\cal D}^C f=0\, .
}

Similarly, the 
fluxes ${\cal F}_{ABC}=E_{CM}\,( {\cal L}_{E_A} E_B{}^M )$ 
and ${\cal F}_A=-e^{2d}\, ({\cal L}_{E_A} e^{-2d})$ with flat indices transform as 
scalars with respect to generalized diffeomorphisms, i.e. 
\eq{
\label{gendiffsflux}
     \delta_\xi {\cal F}_{ABC} = \xi^M \partial_M {\cal F}_{ABC}\,, \qquad
   \delta_\xi {\cal F}_{A}=\xi^M \partial_M {\cal F}_{A}\, .
}
However, under  a local double Lorentz transformation  one gets
as
\eq{
\label{doublelori}
    \delta_{\Lambda} {\cal F}_{ABC}&= 3\, \Big[ {\cal D}_{[A} \Lambda_{BC]}+
      \Lambda_{[A}{}^D {\cal F}_{BC]D}\Big]\,, \qquad
    \delta_{\Lambda} {\cal F}_{A}= {\cal D}^B \Lambda_{BA} + \Lambda_{A}{}^B 
                {\cal F}_B\, ,
}
where the first terms are anomalous. We also write e.g.
$\Delta_{\Lambda} {\cal F}_{ABC}= 3\, {\cal D}_{[A} \Lambda_{BC]}$.
For the relation \eqref{importanta} to be well defined we also require
\eq{    
\label{importantc}
   0=\Delta_{\Lambda}(\Omega^C{}_{AB}\, {\cal D}_C f)=
    ({\cal D}^C \Lambda_{AB}){\cal D}_C f\, ,
}
which could also be read off from \eqref{SCfromClosure}

Moreover, the fluxes  satisfy the  
generalized Bianchi identities
\eq{
\label{bianchi1}
    {\cal D}_{[A} {\cal F}_{BCD]}-{3\over 4} {\cal F}_{[AB}{}^M\, {\cal
          F}_{CD]M}&={\cal Z}_{ABCD}}
and 
\eq{
\label{bianchi2}
    {\cal D}^{M} {\cal F}_{MAB} + 2{\cal D}_{[A} {\cal F}_{B]} -
     {\cal F}^M\, {\cal
          F}_{MAB}&={\cal Z}_{AB}\, ,
}
where the right hand sides are given by 
\eq{
\label{zexpli}
      &{\cal Z}_{ABCD}=-{3\over 4} \Omega_{E[{AB}}\,
    \Omega^E{}_{{CD}]} \\
      &{\cal Z}_{AB}=\Big(\partial^M\partial_M E_{[{A}}{}^N \Big)
    E_{{B}]N}-2\, \Omega^C{}_{AB}\, {\cal D}_C d\, .
}
Both quantities vanish by the strong constraint. As shown
in \cite{Geissbuhler:2013uka}, realizing that
$\Delta_{E_A} {\cal F}_B={\cal Z}_{AB}$ and
$\Delta_{E_A} {\cal F}_{BCD}={\cal Z}_{ABCD}$
this also holds for the closure constraint.

Due to \eqref{gendiffsflux} the DFT action \eqref{dftactionfluxb} is apparently
invariant under generalized diffeomorphisms. Taking the  
anomalous terms in \eqref{doublelori} into account, under local double Lorentz
transformations, the action transforms into a boundary term plus
\eq{ \label{trafoaction}
  \delta_\Lambda  S_{{\rm DFT}}  =  
  \int dX e^{-2d} \Lambda_A{}^C \left( \eta^{AB}-S^{AB}\right) {\cal Z}_{BC}
}
which indeed vanishes for all possible constraints.

The derivative \eqref{deriv} satisfies the commutation relations
\eq{
\label{commalgebra}
[{\cal D}_A,{\cal D}_B]&= {\cal F}^C{}_{AB}\, {\cal D}_C - \Omega^C{}_{AB}\,
{\cal D}_C={\cal F}^C{}_{AB}\, {\cal D}_C
,}
where $\Omega^C{}_{AB}\,{\cal D}_C$
vanishes after invoking either the strong or the closure constraint \eqref{importanta}.

Now, varying the action with respect to the beins, one obtains the
equations of motion
\eq{
\label{dfteom}
   {\cal G}^{[AB]}= {\cal Z}^{AB} + 2 S^{C[A} {\cal D}^{B]} {\cal F}_C +({\cal F}_C-{\cal D}_C)
   \breve{\cal F}^{C[AB]} +\breve{\cal F}^{CD[A}\, {\cal F}_{CD}{}^{B]}=0
}
with
\eq{
   \breve{\cal F}^{ABC}=\breve{S}^{ABCDEF}\,{\cal F}_{DEF} 
    }
and 
\eq{
   \breve{S}^{ABCDEF}=&{1\over 2} S^{AD}\, \eta^{BE}\, \eta^{CF} +
                     {1\over 2} \eta^{AD}\, S^{BE}\, \eta^{CF}+
                     {1\over 2} \eta^{AD}\, \eta^{BE}\, S^{CF}\\
             &-{1\over 2} S^{AD}\, S^{BE}\, S^{CF}\, .
}
Note that the $\Omega$-odd terms in \eqref{dftactionfluxb} do not
contribute to these equations of motion.
The dilaton equation of motion is that the integrand of the action
\eqref{dftactionfluxb} vanishes.
It is remarkable that it is possible  to express the 
equations of motions, including  the gravity 
part, in this  unified way just in terms of doubled fluxes
${\cal F}_{ABC}$ and ${\cal F}_A$.

Finally, let us mention that,
by analyzing a Scherk-Schwarz reduction of DFT, it was pointed 
out in 
\cite{Aldazabal:2011nj,Geissbuhler:2011mx}  
that  the  quadratic constraints of gauged supergravity 
are  satisfied even though the strong constraint is not.
Additionally, in \cite{Grana:2012rr,Geissbuhler:2013uka} 
it was shown that for such Scherk-Schwarz reductions the closure
constraint of DFT is satisfied.
Thus, in a compactified DFT
the strong constraint seems only to be a sufficient but
not a necessary requirement. These Scherk-Schwarz reductions
provide  explicit examples of truly doubled geometries
\cite{Dibitetto:2012rk}.
Whether all such truly non-geometric backgrounds are honest solutions
of string theory is still under debate.

\section{Non-associative deformations of DFT}
\label{sec_main}

In this section  we investigate the generalization of 
the open string analysis from section \ref{sec_open}
to  the closed string, which we 
describe by DFT. As we argued, (on-shell) closed string
scattering amplitudes are not expected to  show
any sign of  non-associativity.
The latter is due to the fact that CFT amplitudes are crossing symmetric,
which correspond to satisfied  Jacobi-identities in an operator formalism.
Therefore, we again expect that the deformation of the effective
action by a (non-associative) tri-product should better be trivial (at least)
on-shell.
However, let us stress that, if one can identify such a specific
non-trivial tri-product, one definitely
has made a big change of the underlying geometry. We will show that, 
under certain conditions, it remarkably 
has no effect for the DFT action. In a similar vein, the conformal
$SL(2,\mathbb C)$ symmetry does not preserve the (radial) ordering
of points on the sphere. Therefore, on-shell one also does not
expect to see any imprint of non-commutativity.

In DFT, there exist  two possible tri-products. First, there is the tri-product 
\eq{
\label{triA}
      f\,\tri\, g\,\tri\, h=f\,g\,h +
      {\ell_s^4\over 6} \breve{\cal F}^{ABC}\, {\cal D}_A f\, {\cal D}_B g\,  {\cal D}_C h 
       +O(\ell_s^8)\, .
}
Since \eqref{triA}  contains the component
 $H^{abc}\, \partial_a f\, \partial_b g\, \partial_c h$, with $H^{ijk}=g^{ii'} g^{jj'} g^{kk'} H_{i'j'k'}$, 
it can be considered  as the DFT generalization of 
the three-product  \eqref{threebracketcon}
with $H$-flux deformation. Even though 
there does not exist 
evidence for the presence of some non-associativity 
for $H$-flux, we study it here, as it is the direct generalization
of the open string story and it still shows some remarkable properties.

The second possibility is the generalization of the tri-product 
with $R^{ijk}$ deformation 
\eq{
\label{triB}
      f\,\tri\, g\,\tri\, h=f\,g\,h +
       {\ell_s^4\over 6} {\cal F}_{ABC}\, {\cal D}^A f\, {\cal D}^B g\,  {\cal D}^C h 
       +O(\ell_s^8) \,. 
}
As mentioned in the introduction, for this case the CFT analysis 
showed some signs of non-associati\-vi\-ty.

In this section we will see that both of these in principle 
possible non-associa\-tive deformations do not lead to
any physical effect in on-shell DFT, though the mechanisms turn out to be 
different for the two cases.

\subsection{A tri-product for $\breve{\cal F}^{ABC}$}
\label{sec_trifhat}

In  analogy to the non-associative product for the open string, we
consider the DFT tri-product
\eq{
\label{trihflux}
      f\,\tri\, g\,\tri\, h=f\,g\,h +
      {\ell_s^4\over 6} \breve{\cal F}^{ABC}\, {\cal D}_A f\, {\cal D}_B g\,  {\cal D}_C h 
       +O\left(\ell_s^8\right)\, .
}
We assume that   $f,g,h$ are scalars under 
generalized diffeomorphisms and are invariant under doubled local Lorentz
transformations.

Invoking the strong or closure constraint, $\breve{\cal F}^{ABC}$ and
 ${\cal D}_A f$ transform as  scalars
under generalized diffeomorphisms so that the tri-product is invariant
under the latter.
The anomalous transformation behavior of the tri-product 
under doubled local Lorentz transformations is
\eq{
     \Delta_{\Lambda}\left( \breve{\cal F}^{ABC}\, {\cal D}_A f\, {\cal D}_B g\,
           {\cal D}_C h \right)= 3 S^{[\underline{A}D} {\cal D}^{\underline B}
        \Lambda^{\underline{C}]}{}_D\, {\cal D}_A f\, {\cal D}_B g\,
           {\cal D}_C h 
}
which vanishes directly for the strong constraint and due to
\eqref{importantc} for the closure constraint.

Now consider the effect of the order $\ell_s^4$ term under the integral.
Performing an integration by parts and using that for both  constraints
we have $[{\cal D}_A,{\cal D}_B]={\cal F}^C{}_{AB} {\cal D}_C$,
we find
\eq{
   \int dX e^{-2d}\,&\breve{\cal F}^{ABC}\, {\cal D}_A f\, {\cal D}_B g\,
        {\cal D}_C h=\int dX \, \partial_M (e^{-2d} V^M)+\\
&\int dX e^{-2d}\, \Big[({\cal F}_C-{\cal D}_C)
   \breve{\cal F}^{C[AB]} +\breve{\cal F}^{CD[A}\, {\cal F}_{CD}{}^{B]}\Big]
    f\, {\cal D}_A g\, {\cal D}_B h \, .
}
with
\eq{
    V^M= E_A{}^M\breve{\cal F}^{ABC}\, f\, {\cal D}_B g\,
        {\cal D}_C h\, 
}
transforming as a vector under generalized diffeomorphisms.
Thus, invoking Stokes theorem
this gives a boundary term, which vanishes on well defined compact doubled
geometries patched by generalized diffeomorphisms and double
Lorentz transformations. Here we have also used the relation
\eq{
     \partial_M (E_A{}^M\, e^{-2d} )=-e^{-2d} {\cal F}_A\, .
}
The second term can be written as
\eq{
  \int dX e^{-2d}\, \Big[ {\cal G}^{[AB]}-2 S^{M[A} {\cal D}^{B]} {\cal F}_M\Big]
    f\, {\cal D}_A g\, {\cal D}_B h =0
}
where, due to \eqref{dfteom},  ${\cal G}^{[AB]}$ vanishes  on-shell 
and the second term vanishes 
 for both the strong and, due to
\eqref{importantb}, also for  the closure constraint.
Thus, we conclude that the 
 order $\ell_s^4$ term in the tri-product is  a surface term 
on-shell. In this respect this tri-product is very similar
to the open string story.

\subsubsection*{Matter corrections}

However, these equations of motion receive stringy higher derivative 
corrections, so that the tri-product, i.e.
the coefficient $\breve{\cal F}^{ABC}$,  needs
to be adjusted accordingly. 
Moreover, coupling DFT to extra matter sources, which, in particular, means
{\it any} additional  field contributing to the energy-momentum tensor,
the equations of motion change  to
\eq{
\label{dfteomsource}
   2 S^{C[A} {\cal D}^{B]} {\cal F}_C +({\cal F}_C-{\cal D}_C)
   \breve{\cal F}^{C[AB]} +\breve{\cal F}^{CD[A}\, {\cal F}_{CD}{}^{B]}= {\cal
     T}^{AB}\, .
}
For instance, including the R-R sector \cite{Hohm:2011zr,Hohm:2011dv}, one can put all R-R fields
in the spinor representation of $O(D,D)$ 
\eq{
{\cal G}=\sum_{n} {e^{\phi}\over n!}\, G^{(n)}_{i_1\ldots i_n}\, e_{a_1}{}^{i_1}\ldots 
        e_{a_n}{}^{i_n}\, \Gamma^{a_1\ldots a_n}|0\rangle\, ,
}
where $\Gamma^{a_1\ldots a_n}$ defines the totally anti-symmetrized
product of $n$ $\Gamma$-matrices.
Then, the R-R contribution to the DFT equation of motion is
\eq{
       {\cal T}^{AB}={1\over 4} \overline{\cal G}\, \Gamma^{AB}\,
 {\cal G}\, .
}
In order to still keep the total derivative property, the only thing one can
do is to adjust the tri-product \eqref{trihflux} as
\eq{
\label{energyadjust}
    f\,\tri\, g\,\tri\, h=\ldots+
       {\ell_s^4\over 18} {\cal T}^{AB}\, \Big( f\, {\cal D}_A g\,
       {\cal D}_B h +  {\cal D}_A f\, {\cal D}_B g\, h+
          {\cal D}_B f\, g\,
       {\cal D}_A h\Big) \, .
}
This means that one already has to introduce a non-trivial two-product
as
\eq{
    f\,\tri_2\, g\,=f\cdot g+
       {\ell_s^4\over 18} {\cal T}^{AB}\, {\cal D}_A f\, {\cal D}_B g 
   +O\left(\ell_s^8\right)\, .
}
Let us discuss its effect for the case that one imposes the strong constraint.
Below the integral the order $\ell_s^4$ correction to this two-product
can be written as
\eq{
\label{matterinter}
   \int dX e^{-2d}\,&{\cal T}^{AB}\, {\cal D}_A f\, {\cal D}_B g\,
        =\int dX \partial_M (\ldots)^M+\\
&\int dX e^{-2d}\, \Big[({\cal F}_A-{\cal D}_A)
   {\cal T}^{AB} -{1\over 2}{\cal T}^{CD}\, {\cal F}_{CD}{}^{B}\Big]
    f\, {\cal D}_B g \, .
}
Employing the Bianchi identities \eqref{bianchi1} and \eqref{bianchi2}
and the strong or the closure constraint, from 
\eqref{dfteomsource} we derive  the
continuity equation for the energy-momentum tensor
\eq{
         ({\cal D}_A -{\cal F}_A) {\cal T}^{AB} + {1\over 2}{\cal F}_{CD}{}^B\, {\cal
    T}^{CD} =S^{CA} \, {\cal D}^B\Big( {\cal D}_A {\cal F}_C-{1\over 2}
        {\cal F}_A\, {\cal F}_C  \Big)\, .
}
Thus, due to the strong constraint the second line in \eqref{matterinter}
vanishes and the order $\ell_s^4$ correction to the two-product
gives a total derivative below the integral.
Note that such a two-product implies 
a two-bracket 
\eq{
[x^i,x^j]={\ell_s^4\over 9} {\cal T}^{ij}\, .
}

Thus, we conclude that, 
due to higher order and source term corrections to the  equations of
motion, the tri-product  needs to  be adjusted accordingly.
For the matter source term, we showed explicitly that at order
$\ell_s^4$ this is indeed possible.
We find it compelling  that the definition of a tri-product
and the DFT/string  equations of motion are related  in this intricate
manner. Deforming the underlying geometry in this non-associative way
does not effect the on-shell DFT.

\subsection{A tri-product for ${\cal F}_{ABC}$}

Now consider the DFT generalization of
the tri-product \eqref{threebracketcon} 
\eq{
\label{triproda}
      f\,\tri\, g\,\tri\, h=f\,g\,h +
       {\ell_s^4\over 6} {\cal F}_{ABC}\, {\cal D}^A f\, {\cal D}^B g\,  {\cal D}^C h 
       +O(\ell_s^8)\, .
}
Note that, once the strong or closure constraint is imposed, the order $\ell_s^4$
term in \eqref{triproda} transforms as a scalar under 
generalized diffeomorphisms  if $f,g,h$ are scalars.
In addition this  tri-product is also invariant under  
local double  Lorentz transformations.
However, a second look reveals that
this is trivial as, imposing either constraint, one
immediately realizes that due to \eqref{importanta} 
the whole order $\ell_s^4$ term actually
{\it vanishes}. 
Thus, in this constrained DFT framework  this tri-product is
actually trivial.

For illustrative purposes, nevertheless  let us
apply a partial integration to the tri-product 
\eqref{triproda} written below an integral.
The order $\ell_s^4$ term can be written as
\eq{
\label{werder}
   \int dX e^{-2d}\,{\cal F}_{ABC}\, {\cal D}^A f\, &{\cal D}^B g\,
        {\cal D}^C h=\int dX \,\partial^M (\ldots)_M-\\
&\int dX e^{-2d}\, \Big[({\cal D}^C-{\cal F}^C)
   {\cal F}_{CAB} \Big]
   {\cal D}^A f\, {\cal D}^B g\, h 
}
where the term in the last line can be written as
\eq{
 \int &dX e^{-2d}\, \Big[ {\cal Z}_{AB}-2 {\cal D}_{[A} {\cal F}_{B]} \Big]
   {\cal D}^A f\, {\cal D}^B g\, h \, .
}
Here we have used ${\cal F}_{MN[A} {\cal F}^{MN}{}_{B]}=0$.
Consistently, due to the Bianchi-identity \eqref{bianchi2}
and the relation \eqref{importantb} this expression vanishes
for both constraints.
Since the terms appearing in this computation are related to
the ones appearing in a topological Bianchi identity and not a dynamical 
equation of motion, one might expect that 
there are no stringy higher order derivative  corrections to the, in general,
non-constant  tri-product  parameter ${\cal F}_{ABC}$.

\subsubsection*{Comments on relaxing the closure constraint}

Relaxing even the closure constraint is the only option to get a non-trivial 
tri-product \eqref{triproda}. For compact configurations it is clear
that string theory contains momentum and winding modes
not subject to the weak and consequently the strong constraint.
For instance, for a toroidal compactification, the level matching 
condition becomes
\eq{
\label{levelmatch}
         L_0-\ov L_0 =  \alpha'\, p\cdot w + N-\ov N  =0
}
where $N$ and $\ov N$ denote the number of  left and right-moving oscillator
excitations. Including these modes is expected to go beyond the realm
of DFT. 

Another way of relaxing the closure constraint could be  by
splitting the fluxes into backgrounds and fluctuations and relaxing 
the strong and closure constraint between the two. Whether this is  
an allowed relaxation in DFT remains to be seen and is beyond
the scope of this paper. Here we just discuss its consequences
for the tri-product.

Independent of how actually the constraints are relaxed, let us now 
 discuss the consequences for the tri-product.
Up to boundary terms, after partially integrating 
the order $\ell_s^4$ term under the integral we get
\eq{ \label{triproductafterPI}
  \int dX e^{-2d}\, \Big[({\cal D}^C-{\cal F}^C)
   {\cal F}_{C[AB]} +2\,{\Omega}_{CD[A}\, {\cal F}_{B]}{}^{CD}\Big]
    ({\cal D}^A f)\, ({\cal D}^B g)\, h \, .
}
The additional term compared to \eqref{werder} arises from 
the $\Omega$ term in the commutator \eqref{commalgebra} when violating closure.
Taking into account that, in string theory, non-associativity should still be
vanishing at least on shell, we can imagine two ways to proceed from here.

First, we can require a new constraint
\eq{
\label{newstrong}
        \zeta_{AB} \,{\cal D}^A f\, {\cal D}^B g =0
}
with
\eq{
    \zeta_{AB}= ({\cal D}^C-{\cal F}^C)
   {\cal F}_{C[AB]} +2\,{\Omega}_{CD[A}\, {\cal F}_{B]}{}^{CD}
\, }
that is weaker than the closure constraint.
The second possibility is to cancel these terms by an appropriately adjusted
tri-product
\eq{
\label{triadjust4}
 f\,\tri\, g\,\tri\, h=f\,g\,h &+ 
       {\ell_s^4\over 6}\, \, {\cal F}_{ABC}\, \,{\cal D}^A f\, {\cal D}^B g\,  {\cal D}^C h \\
       &+{\ell_s^4\over 18} \, \, \zeta_{AB}\, \, \Big( f\, {\cal D}^A g\,
       {\cal D}^B h +  {\cal D}^A f\, {\cal D}^B g\, h+
          {\cal D}^B f\, g\,
       {\cal D}^A h\Big) \, .
}
Note that one can rewrite the adjusted tri-product \eqref{triadjust4} as
\eq{
\label{triadjust2}
 f\,\tri\, g\,\tri\, h=f\,g\,h &+ \, e^{2d}\,  \partial_M \left({\ell_s^4\over 6} E_A{}^M e^{-2d} 
        {\cal F}_{ABC}\,  f\, {\cal D}^B g\,  {\cal D}^C h + cycl_{f, g, h} \right)
}
showing that it is really designed to give a boundary term below
the integral. One can show that also the induced two-product  
gives a boundary term if written under an integral.

Summarizing, relaxing the closure constraint, 
one can either impose \eqref{newstrong} or define the tri-product
deformation trivially as a total derivative. In both cases one formally has non-vanishing 
brackets \eqref{commalg} and \eqref{commalggen} that leave no trace under an action integral.

\subsubsection*{Holonomic basis}

In order to see more concretely what is happening here, let us consider
as an example a holonomic basis with  $B_{ab}=0$, $f_{ab}{}^c=0$ and $\tilde f^{ab}{}_c=0$.
In this case one finds
\eq{ {\cal F}_{ABC}\, {\cal D}^A f\, &{\cal D}^B g\,  {\cal D}^C h =
      R^{ijk}\, \partial_i f \, \partial_j g\,  \partial_k h +\\
   &\phantom{aaaaaaaaaaaaaa}
     Q_{k}{}^{ij} \,\Big( \partial_i f \, \partial_j g \,\big(\tilde\partial^k + \beta^{kl}
      \partial_l\big)h + {\rm cycl}_{f,g,h} \Big)\\
  =\, &3\Big(\tilde\partial^{[\underline{i}} \beta^{\underline{jk}]} +
     \beta^{[\underline{i}m} {\partial}_m \beta^{\underline{jk}]} \Big)
   \partial_i f \, \partial_j g\,  \partial_k h \\
  &-3\Big(
     \beta^{[\underline{i}m} {\partial}_m \beta^{\underline{jk}]} \Big)
   \partial_i f \, \partial_j g\,  \partial_k h+
    \partial_{k}\beta^{ij} 
\,\Big( \partial_i f \, \partial_j g \,\tilde\partial^k h+
   {\rm cycl}_{f,g,h}\Big)
}
where we have split the $R$-flux as
\eq{
\label{splitrflux}
 R^{ijk} = \hat R^{ijk} + R^{ijk}_{\rm cl}= 3\Big(\tilde\partial^{[\underline{i}} \beta^{\underline{jk}]} +
     \beta^{[\underline{i}m} {\partial}_m \beta^{\underline{jk}]} \Big)\, .
}
Therefore, the second and third term cancel and the sum of the first and fourth
vanish by the constraint. In particular, this means that in DFT
the classical part $R^{ijk}_{\rm cl}=\beta^{[\underline{i}m} {\partial}_m
    \beta^{\underline{jk}]}$  does not contribute to the tri-product.

In order to derive the tri-bracket among three coordinates, let us choose  for the three functions
$f=x^i$, $g=x^j$ and $h=x^k$.
Without imposing neither  the strong nor the closure   constraint\footnote{The CFT computations performed in
  \cite{Blumenhagen:2010hj,Lust:2010iy,Blumenhagen:2011ph,Condeescu:2012sp,Andriot:2012vb}
  were not imposing any
constraints so that they can be considered to be reliable
for the compact torus case for which the level matching condition is 
\eqref{levelmatch}}
the resulting tri-bracket is then  given by
\eq{
       [x^i,x^j,x^k]=\ell_s^4\,  \hat R^{ijk}\, ,
}
and, in particular,  only contains the $R$-flux   $\hat R^{ijk}$.

Let us also consider the general commutator
\eqref{commalggen} for the case that both $Q$- and $R$-flux is present
in more detail.
Our DFT analysis suggests that the commutator for general functions
should be defined as
\eq{
           -{3i\hbar\over \ell_s^4}\,    [f,g]=R^{ijk} &\ \partial_i f\, \partial_j g \,\partial_k
              +  Q_k{}^{ij} \Big( \partial_i f\, \partial_j g \,
                 \big(  {\tilde \partial}^k + \beta^{kl} \partial_l \big) +\\[0.1cm]
   &     \big(  {\tilde \partial}^k + \beta^{kl} \partial_l \big)f\, 
       \partial_i g\,
        \partial_j  +
    \partial_j f\,  
                 \big(  {\tilde \partial}^k + \beta^{kl} \partial_l \big)g\, 
        \partial_i \Big)\, .
}
Inserting the definition of the $R$-flux \eqref{splitrflux}, again
the term $R^{ijk}_{\rm cl}$ completely cancels against terms appearing
in the $Q$-flux contribution and we are left with
\eq{
-{3i\hbar\over \ell_s^4}\, [f,g]=\hat R^{ijk} &\ \partial_i f\, \partial_j g\, \partial_k
              +\\
  &  Q_k{}^{ij} \Big( \partial_i f\, \partial_j g \,
                 {\tilde \partial}^k + 
   {\tilde \partial}^k f\,  \partial_i g\, \partial_j  +
    \partial_j f\,  {\tilde \partial}^k g\,  \partial_i \Big)\, .
}
Note that, invoking the constraint, the commutator vanishes.
Computing the commutation relations for the coordinate
functions, without imposing any constraint, one finds
\eq{
\label{commres}
      [x^i,x^j]&=i{\ell_s^4\over 3\hbar} \, \Big( \hat R^{ijk} \partial_k + Q_k{}^{ij} \, {\tilde \partial}^k \Big)\, ,\\[0.1cm]
      [x^i,\tilde x_k]&= -i{\ell_s^4\over 3\hbar}\, Q_k{}^{ij} \partial_j\, .
 }
Thus,  DFT suggests that the interpretation of
the commutation relation \eqref{commalggen} in terms of derivatives 
is \eqref{commres}.
In particular, the contribution $R^{ijk}_{\rm cl}$ drops out
and all  commutators vanish after imposing any constraint.

\subsubsection*{Higher order corrections}

At leading order in derivatives of ${\cal F}_{ABC}$
there is a natural candidate for the all order in $\ell_s^4$
tri-product, namely
\eq{
\label{tri-product-SS}
   (f\,\tri\, g\, \tri\, h)({X}) = \exp\biggl(
   {\ell_s^4\over 6}\, {\cal F}_{ABC}\,
      {\cal D}_{{X}_1}^{A}\,{\cal D}_{{X}_2}^{B}\,{\cal D}_{{X}_3}^{C} \biggr)\, f({X}_1)\, g({X}_2)\,
   h({X}_3)\Bigr|_{{X}} \;.
}
At leading order in $({\cal D} {\cal F}_{ABC})$, 
except $fgh$, all terms  
give a total derivative below the integral.
The appearing derivatives can be canceled by defining
the overall tri-product as
\eq{
      f\,\blacktri\, g\,\blacktri\, &h=f\,\tri\,  g\, \tri\, h +
     \sum_{k=2}^\infty
{\ell_s^{4k}\over 3\, 6^k k!} \bigg\{ {\cal F}_{A_1 B_1 D}
 \, {\cal D}^{D}\left(
   {\cal F}_{A_2 B_2 C_2}\ldots   {\cal F}_{A_k B_k C_k}\right)\\
 &  \Big( 
    ({\cal D}^{A_1}\ldots {\cal D}^{A_k} f)
    ({\cal D}^{B_1}\ldots {\cal D}^{B_k} g)
    ({\cal D}^{C_2}\ldots {\cal D}^{C_k} h)+{\rm cycl}_{\{f,g,h\}}\Big) 
\bigg\}\, .
}
This product is designed to satisfy
\eq{
    \int dX\, e^{-2d}\, f\,\blacktri\, g\,\blacktri\, h=
     \int dX\, e^{-2d} f\, g\, h\, .
}
A possible generalization of the tri-product to the product
of $K$ functions is presented in the appendix.

\section{Conclusions}

Using the flux formulation of DFT,
we have analyzed the consequences of introducing non-associativity via
a  non-trivial tri-product for the functions on the manifold.
We analyzed two different such non-associative deformations.
For the first the deforming flux was given by $\breve{\cal F}^{ABC}$ 
and for the second by ${\cal F}_{ABC}$.
The first case is the DFT generalization of the $H^{ijk}$-flux deformation
and the second one the generalization of the $R^{ijk}$-flux 
deformation. 

We argued  from conformal field theory that 
on-shell any non-associative deformation should
not lead to any physical effect.
Note that in the open string case, the situation is different.
There the DBI action can be 
expressed in the Seiberg-Witten limit  as a non-commutative gauge theory
and  the higher orders in the star-product 
really contribute physical  terms to the  deformed action. However,
also here cyclicity and associativity are preserved on-shell.

The $\breve{\cal F}^{ABC}$ flux case is conceptually very close
to its open string analogue. Similarly, we found that, 
at leading order in $\ell_s^4$, the deformation 
gives a boundary term
under the integral if the DFT equations of motion are satisfied and
the strong or closure constraint is employed. 
We showed that, for additional matter
contributions, the tri-product can be adjusted accordingly. This led
to a new deformation of the two-product, whose on-shell triviality
was guaranteed by the continuity equation of the energy momentum tensor. 
This means that on-shell DFT or string theory cannot distinguish
between on ordinary smooth geometry and a fuzzy one with 
fundamental tri-bracket
\eq{
       [x^i,x^j,x^k]=\ell_s^4\,  H^{ijk}\, .
}
Even though  from
\cite{Blumenhagen:2010hj,Lust:2010iy,Blumenhagen:2011ph,Condeescu:2012sp,Andriot:2012vb}  
we do not have any evidence for such a non-associative
behavior of the coordinates, we find this a remarkable property
of DFT. Turning the logic around, up to the dilaton sector, 
one can  derive the DFT equations of motion from the concept
of the {\it absence of on-shell non-associativity}. 
We emphasize, that in the flux formulation of DFT also the gravity
part is fully encoded in the generalized three-form flux.
At least in spirit, this is very similar to the familiar magnetic monopole
example discussed in the first section.

The ${\cal F}_{ABC}$ flux case is the one where non-associativity  was
expected. We realized that in the DFT  framework this tri-product actually
vanishes
after  imposing  either the strong or the closure constraint. 
Therefore, in order to get something non-trivial even the closure constraint
need to be weakened.
Only then one  could obtain a non-associative deformation of the target space action
with the three-bracket for the internal coordinates $x^i$ being 
\eq{
       [x^i,x^j,x^k]=\ell_s^4\, \hat R^{ijk} \, .
}
Again note that  the  
$\hat R^{ijk}$ only contain the winding part of the full $R$-flux, the
classical part has canceled  out. 

On a more speculative level, we also proposed a generalization of the
tri-product to higher orders in $\ell_s^4$ and for products
of $K$-terms.

Summarizing, the resolution to the initially raised paradox is 
that one can have 
a non-associative deformation of the target space, while nothing of
it is immediately apparent in the effective 
string and DFT actions for the massless modes.
Deforming the product to a tri-product we have found two different
ways how such a deformation can become trivial (on-shell).

\begin{figure}[ht]
  \centering
  \includegraphics[width=0.70\textwidth]{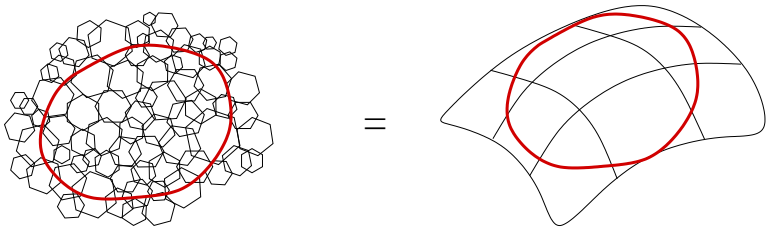}
  \begin{picture}(0,0)
   
  \end{picture}
  \caption{\small Stringy equivalence between fuzzy non-associative geometry 
      and smooth Riemannian geometry.}
  \label{fig:nca}
\end{figure}

One could  imagine that, due to the finite
size and resolution of the string, there exists a certain non-associative
deformation of the target space that is ``under the radar''
of the string. Therefore, string theory can very well
admit such non-geometric space as honest backgrounds.
An artist's  impression of this picture
is presented in figure \ref{fig:nca}.

It would be interesting to carry out a similar analysis
for the (precursor) non-commutative closed string star product defined on phase 
space, which was introduced and discussed in 
\cite{Mylonas:2012pg,Bakas:2013jwa}.
Moreover, one could   contemplate what other deeper 
conceptual consequences 
the existence of such a  non-geometric regime of string theory might have. 
Including also the massive string states, 
can it be generalized to string field theory?
Does there exist an analogous structure for M-theory?

\vspace{0.5cm}

\noindent
\subsubsection*{Acknowledgments}
We would like to thank Ioannis Bakas, David Berenstein, Andreas Deser, Daniela Herschmann, Erik Plauschinn, Felix Rennecke,
Christian Schmid, Peter Schupp and Richard Szabo   for discussion. 
We also acknowledge 
that this project was strongly influenced by
the nice atmosphere at the
{\it Workshop on  Noncommutative Field Theory and Gravity} at the
Corfu Summer Institute 2013. 
This work was partially supported by the ERC Advanced Grant "Strings and Gravity"
(Grant.No. 32004) and by the DFG cluster of excellence "Origin and Structure of the Universe".

\clearpage
\appendix


\section{$K$ tri-product}
\label{app_b}

In this appendix we discuss how to treat terms which involve for instance
a product of $K$ functions.
Clearly, e.g. for $K=4$ this is not defined by
an iteration of the tri-product \eqref{tri-product-SS}.
{}From the analysis of multiple tachyon scattering amplitudes in CFT, 
in \cite{Blumenhagen:2011ph} 
a proposal was made, how to deform the product of $K$ functions.
Analogously, at leading order in $({\cal D} {\cal F}_{ABC})$ (or
$({\cal D} \breve{\cal F}_{ABC})$)
we now  define the $K$-fold  tri-product as
\begin{eqnarray}
\label{ntriprod}
   &&\!\!\!\!(f_1\, \tri_K\,  f_2\, \tri_K \ldots \tri_K\,  f_K)({X}) \stackrel{\rm def}{=} \\
   &&\phantom{aaaaaa}\exp\biggl( {{\ell_s^4\over 6}}\, {\cal F}_{ABC}\,\!\!\!\!\! \sum_{1\le a< b < c\le K}
     \!\!\!\!  \, 
      {\cal D}_{{X}_a}^{A}\,{\cal D}_{{X}_b}^{B} {\cal D}_{{X}_c}^{C} \biggr)\, 
   f_1({X}_1)\, f_2({X}_2)\ldots
   f_K({X}_K)\Bigr|_{{X}} \;.\nonumber
\end{eqnarray}
Below  we prove the remarkable feature that for each $K$
all terms beyond leading order  give
a total derivative under the internal integral, i.e.
\eq{
    \int d{X} \; e^{-2d}\, f_1\,\tri_K\, f_2\,\tri_K\, \ldots
    \tri_K\, f_K=
     \int d{X}\; e^{-2d} f_1\, f_2\, \dots \, f_K \, .
 }
Moreover, this $K$ tri-product has the property
\eq{  f_1\,\tri_K\, \ldots
    \tri_K\, 1= f_1\,\tri_{K-1}\, \ldots
    \tri_{K-1}\, f_{K-1}
}     
which suggests  to define $f_1\tri_2 f_2=f_1\cdot f_2$, i.e. the
two tri-product is the ordinary multiplication of functions.

Note that the total derivative  property does {\it not} hold for  a similar
definition of an $K$ star-product
\begin{eqnarray}
\label{nstarprod}
   &&\!\!\!\!(f_1\, \star_K\,  f_2\, \star_K \ldots \star_K\,  f_K)({X}) \stackrel{\rm def}{=} \\
   &&\phantom{aaaaaa}\exp\biggl( i\, {{\ell_s^2\over 2}}\, \theta^{IJ}\,\!\!\!\!\! \sum_{1\le a< b \le K}
     \!\!\!\!  \, 
      \partial^{{X}_a}_{I}\,\partial^{{X}_b}_{J}  \biggr)\, 
   f_1({X}_1)\, f_2({X}_2)\ldots
   f_K({X}_K)\Bigr|_{{X}} \;,\nonumber
\end{eqnarray}
This is why for the open string case, the non-commutativity of the 
underlying space-time has a non-trivial effect on the action.

\subsubsection*{Proof}

Here we present the proof that at leading order
in ${\cal D} {\cal F}_{ABC}$  
the  $K$ tri-product \eqref{ntriprod}
gives a total derivative under the integral, i.e.
\eq{
    \int d{X} \; e^{-2d}\, f_1\,\tri_K\, \ldots \,
    \tri_K\, f_K=
     \int d{X}\; e^{-2d} f_1\, \dots \, f_K \, .
 }
We first consider just the order $\ell_s^4$ term, which is given by
\eq{
{{\ell_s^4\over 6}}\, {\cal F}_{ABC}\,\!\!\!\!\! \sum_{1\le a< b < c\le K}
     \!\!\!\!  \, 
      {\cal D}_{{X}_a}^{A}\,{\cal D}_{{X}_b}^{B} {\cal D}_{{X}_c}^{C} \biggr(\, 
   f_1({X}_1)\, f_2({X}_2)\ldots
   f_K({X}_K) \biggr) \Bigr|_{{X}} .\;\nonumber
}
Inspection reveals, that the ${ K \choose 3}$ terms can be grouped together 
as
\eq{
  \, &{\cal D}^A (f_1) \; {\cal D}^B f_2 \; {\cal D}^C (f_3 \dots f_{K}) \\
  + \,&{\cal D}^A (f_1 f_2)\; {\cal D}^B f_3 \; {\cal D}^C (f_4 \dots f_{K}) \\
  + \, &{\cal D}^A (f_1 f_2 f_3)\;  {\cal D}^B f_4\;  {\cal D}^C (f_5 \dots f_{K}) \\
  + \, & \dots \\
  + \, & {\cal D}^A (f_1 \dots f_{K-2}) \; {\cal D}^B f_{K-1} \; {\cal D}^C
  (f_K)\, .
}
Note that the sum fixes the 
order of the derivatives and the number of terms is correct, since
\eq{
{K \choose 3} = 1\cdot (K-2) + 2 \cdot (K-3) + \dots + (K-2)\cdot 1.
}
As one can see, the $K$ tri-product splits into $K-2$ three tri-products and 
therefore shares its properties under an integral. 
The higher order terms follow immediately
by iteration. This is owed to the fact that, 
in the derivation of the total derivative 
property, only first three derivatives are relevant.




\clearpage
\bibliography{references}  
\bibliographystyle{utphys}


\end{document}